\documentclass[aps,prb,floatfix,amsmath,amssymb,superscriptaddress,twocolumn,nofootinbib]{revtex4-2}
\usepackage{bm}                     
\usepackage{appendix}
\usepackage[latin1]{inputenc}
\usepackage{dcolumn}      
\usepackage{bbm}
\usepackage{color}
\usepackage{xcolor}
\usepackage[caption=false]{subfig}

\usepackage{amsmath}
\usepackage[mathscr]{eucal}
\usepackage{latexsym}
\usepackage{amsbsy}
\usepackage{float}
\usepackage{graphicx}
\usepackage{epsfig}
\usepackage{wrapfig}
\usepackage{epsf}
\usepackage{float}
\usepackage[normalem]{ulem}
\usepackage{qcircuit}

\definecolor{darkblue}{HTML}{004D6B}
\definecolor{darkred}{HTML}{8c1515}
\definecolor{darkgreen}{HTML}{006400}
\usepackage{hyperref}
\hypersetup{ colorlinks=true, urlcolor=darkblue, citecolor=darkred,
    linkcolor=darkblue, breaklinks }
\usepackage{parskip}

\newcommand{\be}{\begin{equation}}
\newcommand{\ee}{\end{equation}}
\newcommand{\ba}{\begin{array}{l}}
\newcommand{\ea}{\end{array}}
\newcommand{\re}[1]{(\ref{#1})}
\newcommand{\ci}[1]{\cite{#1}}
\newcommand{\banonum}{\begin{eqnarray*}}
\newcommand{\eanonum}{\end{eqnarray*}}
\newcommand{\baa}{\begin{eqnarray}}
\newcommand{\eaa}{\end{eqnarray}}
\newcommand{\bfr}{\begin{flushright}}
\newcommand{\efr}{\end{flushright}}
\newcommand{\bfl}{\begin{flushleft}}
\newcommand{\efl}{\end{flushleft}}
\newcommand{\lab}[1]{\label{#1}}

\begin{document}

\title{Optical high harmonic generation in Dirac materials}
\author{S. Rakhmanov}
\affiliation{Chirchik State Pedagogical University, 104 Amur Temur Str., 111700 Chirchik, Uzbekistan}
\author{K. Matchonov}
\affiliation{National University of Uzbekistan, 4 University Str., 100174 Tashkent, Uzbekistan}
\author{H. Yusupov}
\affiliation{Kimyo International University in Tashkent, 156 Usman Nasyr Str., 100121, Tashkent, Uzbekistan}
\author{K. Nasriddinov}
\affiliation{Chirchik State Pedagogical University, 104 Amur Temur Str., 111700 Chirchik, Uzbekistan}
\author{D. Matrasulov}
\affiliation{Turin Polytechnic University in Tashkent, 17 Niyazov Str., 100095 Tashkent, Uzbekistan}


\begin{abstract} 
We study high-order harmonic generation by optically driven one- and two-dimensional  hydrogen-like atoms formed by Coulomb imurities in graphene.  The time-dependent Dirac equations with Coulomb plus time-periodic monochromatic field potentials are solved for both cases. Such characteristics of the optical high harmonic generation, as average dipole moment and high harmonic generation spectra, are computed. A sketch for table-top experimental realization of the considered models is proposed.
\end{abstract}

\maketitle

\section{Introduction}
Optical high-harmonic generation is a quantum (microscopic) manifestation of the nonlinear optical phenomenon called frequency conversion or harmonic generation. Unlike its macroscopic (nonlinear) counterpart, it may occur in the interaction of single or few atoms (molecules) with an external optical field. The phenomenon is important both, from basic research, as well as practical viewpoints. Fundamental aspect of the effect is related to the fact that can be  very powerful tool for the study of ultrafast (attosecond) quantum phenomena, while the practical one is of direct relevance to the problem of creation ultrashort laser pulses with tunable properties.  An important problem in the study of high harmonic generation is finding the regime that makes the process maximally tunable. Another important issue is its experimental realization in such way that facilitates the practical application of the process in modern quantum and optoelectronic technologies. So far optical high harmonic generation has been mostly studied for nonrelativistic systems (see, e.g., book \ci{Boydbook} and Refs.~\ci{Tong97}-\ci{Rakhmanov23} for review). In this paper, we consider HHG in the so-called Dirac materials caused by the interaction with an external optical field. Dirac materials are defined as the structure where quasiparticles (electron, hole, etc) dynamics are described in terms of the one- or two-dimensional Dirac equation \ci{Brey06}-\ci{Wang13}. The most famous Dirac material is graphene, which is a one-atom thick two-dimensional material. In low-energy regime, quasiparticles in graphene is described in terms of the Dirac equation. When graphene contains charged(Coulomb) impurities, they form two-dimensional atoms, which can "mimic" a heavy relativistic atom \ci{Wang13,Wang12,Pereira07}. Different aspects of planar atoms formed by Coulomb impurities are studied in Refs.~\ci{Mayer14}-\ci{Rakhmanov24_01}. Carbon nanotubes, graphene nanoribbons and topological materials can be also considered as Dirac materials. Here we consider high harmonic generation by Coulomb impurities in graphene in their interaction with external linearly polarized optical field. The system atom+optical field is described in terms of the time-dependent Dirac equation with Coulomb plus monochromatic field potential. Two special cases are considered: Planar atom formed in  bulk graphene monolayer and 1D atom formed in graphene nanoribbon. The advantage of considering HHG in Dirac materials comes from the fact that such models can "low-cost"  experimental realization, i.e. at the table-top level. In addition, being low-dimensional systems, they have more tools for tuning the HHG process compared to 3D models.  The paper is organized as follows. In the next section we briefly recall one- and two- dimensional relativistic atoms formed by Coulomb impurities in graphene nanoribbon and bulk graphene sheet, respectively. Section III presents description of the interaction of one-dimensional and planar relativistic atoms with external monochromatic field. Section IV presents detailed study of  high harmonic generation by the above atoms. In section V a sketch for experimental realization is discussed. Finally, section VI presents some concluding remarks. 


\section{Relativistic atoms formed by Coulomb impurities in graphene}
When a charge, i.e. Coulomb impurity is doped in graphene it can capture an electron and planar relativistic atom, where electrons are described in terms of the 2D Dirac equation with Coulomb potential \ci{Wang13,Wang12,Pereira07}.
Properties such "atoms" have been earlier extensively studied. Dirac equation based description of the electronic structure and  scattering was considered in detail in the Ref.~\ci{Novikov}. The first study of the planar Dirac equation for Coulomb potential, including the supercritical states dates back to the series of papers by Khalilov, et. al  \ci{Khalilov}. The 2D  the Dirac equation based study of the  Coulomb impurities in grahene was considered first in  the Ref.~\ci{Biswas07,Pereira08}. Later the study was extended to the case of critical and supercritical impurities.
In theory of relativistic atoms the "supercritical states" mean the electronic states when the energy level of a relativistic atom reached or embedded into the the Dirac sea.  Basic mechanisms of vacuum effects including quasiparticle pair creation in graphene induced by in time-dependent electromagnetic field are studied the Refs.~\ci{Allor}-\ci{Fillion}.

\subsection{One-dimensional atom in graphene nanoribbon}

Artificial, relativistic 1D atoms can be formed by Coulomb impurities doped into  gapped graphene nanoribbon. In such case the electron motion is described in terms of the one-dimensional massive Dirac equation.
Such 1D atoms was considered in detail in the Ref.~\ci{Downing14}. Here, following the Ref.~\ci{Downing14},  we briefly recall the Dirac equation for one-dmensional Coulomb potential 
Electron motion in 1D (shifted) Coulomb potential can be described in terms of the following Dirac Hamiltonian \cite{Downing14} ($v_F=\hbar=M=1$):

\begin{equation}
    \hat{H}=\left(\begin{array}{cc}0 & -i \partial_x-i \\ -i \partial_x+i & 0 \end{array}\right)-\frac{\alpha}{a+|x|}, \label{hamil}
\end{equation}
where $a-$ shift length, the dimensionless number $\alpha$ 
parametrizes the charge of the Coulomb impurity (multiplied by the 
fine structure constant and the inverse dielectric constant). 
A wave function of the one-dimensional relativistic hydrogen-like atom $\psi(x)=\left(\begin{array}{c} \psi_{1}(x) \\ \psi_{2}(x)  \end{array}\right) $ is found from the Dirac equation given by
\begin{equation}
    \hat{H}\psi(x)=E\psi(x), \label{eq01}
\end{equation}
here $E-$ eigenvalues.

Using the  unitary transform \cite{Downing14}
$$U=\frac{1}{\sqrt{2}}\left(\begin{array}{cc} 1 & 1 \\ 1 & -1  \end{array}\right),$$ 
for Hamiltonian \re{hamil}, one can be written as

\begin{equation}
    \left(\begin{array}{cc} -i \partial_x & i \\ -i & i \partial_x \end{array}\right) \left(\begin{array}{c} \psi_{1}(x) \\ \psi_{2}(x)  \end{array}\right)= \left(E+\frac{\alpha}{a+|x|}\right)\left(\begin{array}{c} \psi_{1}(x) \\ \psi_{2}(x)  \end{array}\right). \label{eq02}
\end{equation}
 
Furthermore, the solution of Eq.~\re{eq02} can be defined in two regions. For region I belonging to  $x<0$ \cite{Downing14} we have
\begin{equation}
    \psi_I=\frac{c_I}{\sqrt{a}}\left(\begin{array}{c}(\kappa+iE)W_{\mu,\nu+1}(2\kappa(a-x))\\W_{\mu,\nu}(2\kappa(a-x))\end{array}\right). \label{sol01}
\end{equation}

For the region II, where $x>0$ one can write the solution as \cite{Downing14}

\begin{equation}
    \psi_{II}=\frac{c_{II}}{\sqrt{a}}\left(\begin{array}{c}W_{\mu,\nu}(2\kappa(a+x))\\-(\kappa+iE)W_{\mu,\nu+1}(2\kappa(a+x))\end{array}\right), \label{sol02}
\end{equation}
where 
\begin{equation}
    W_{\mu,\nu}(\xi)=\xi^{1/2+\nu}e^{-\xi/2}U(\frac{1}{2}+\nu-\mu,1+2\nu,\xi)
\end{equation}
is the Whittaker function with $U(\alpha,\beta,\xi)$ which can be written in terms of the confluent hypergeometric functions as \cite{Abramowitz}
\begin{equation}
\begin{array}{ll}
     U(\alpha,\beta,\xi)=  \\
     \\
     \frac{\Gamma(1-\beta)}{\Gamma(\alpha-\beta+1)}F(\alpha,\beta,\xi)+\frac{\Gamma(\beta-1)}{\Gamma(\alpha)}F(\alpha-\beta+1,2-\beta,\xi) 
\end{array}
\end{equation}
with $\kappa=\sqrt{1-E^2}$, $\mu=\frac{E\alpha}{\kappa}$, $\nu=i\alpha-\frac{1}{2}$.
The normalization constants, $c_{I}\;$ and $c_{II}$ can be found from the normalization conditions written in terms of the components of spinor $\psi$, while from continuity of the wave function at the boundary of domains one can obtain a secular equation for finding the eigenvalues, $E_n$ \cite{Downing14}.
\subsection{Planar relativistic atom in monolayer graphene}
Planar relativistic atoms can be formed when by Coulomb impurities in monolayer graphene sheet. Electronic properties of such atom are studied in detail by Novikov in the Ref.~\cite{Novikov}. Here we briefly recall basic points of that study following the Ref.~\cite{Novikov}. 

Electron motion of mass M dynamics in gapped graphene in the presence of Coulomb charge is described  by 2D Dirac Hamiltonian ($\hbar =v_F= 1$):
\be
H_0 = \sigma_x p_x +\sigma_y p_y+M\sigma_z -\frac{\alpha}{r} = \left(\begin{array}{cc} M-\frac{\alpha}{r} & p_x -ip_y\\p_x+ip_y & -M-\frac{\alpha}{r}
\end{array} \right) ,
\lab{hamilt}
\ee where $r=\sqrt{x^2+y^2}$, $p_x =-i\frac{\partial}{\partial x}, p_y =-i\frac{\partial}{\partial y},$ $\alpha$ is the charge if the impurity (multiplied by fine structure and dielectric constants), $\sigma_x$, $\sigma_y$ and $\sigma_z$ are the Pauli matrices given as
\be \sigma_x=\left(\begin{array}{cc} 0 & 1\\1 & 0
\end{array} \right), \;\;\sigma_y=\left(\begin{array}{cc} 0 & i\\-i & 0
\end{array} \right),  \;\;\sigma_z=\left(\begin{array}{cc} 1 & 0\\0 & -1
\end{array} \right).\ee

In the absence of time-dependent interaction the wave function of quasiparticles obeys the stationary Dirac equation given by
\be
 H_0\psi =E\psi,
\lab{Dir2}
\ee
where
\begin{equation}
\psi(x,y) = \left\{  \begin{array}{c} \phi  \\
\\\chi \ \end{array} \right\}.
\label{WF1}
\end{equation}

Angular and radial variables can be separated as
\be
\psi_j(\vec
r)=\left(\begin{array}{ccc}F(r)\Phi_{j-1/2}(\theta)\\iG(r)\Phi_{j+1/2}(\theta)\end{array}\right) \lab{wf000}
\ee
with 
\be  \Phi_m(\theta)=\frac{1}{\sqrt{2\pi}}e^{im\theta} \lab{angwf} \ee
and for discrete spectrum, $|E|<M$ we have the wave functions ($\hbar=v_F=1$) \ci{Novikov}
\begin{equation}
\begin{split}
\left(\begin{array}{c}F\\G\end{array}\right)=\frac{(-1)^n\lambda^{3/2}}{M\Gamma(1+2\gamma)}
\sqrt{\frac{\Gamma(1+2\gamma+n)(M \pm
E)}{(j+\alpha/\lambda)\alpha n!}} \times \\
e^{-\lambda r}(2\lambda
r)^{\gamma-1/2}\bigg(\left(j+M\frac{\alpha}{\lambda}\right)F(-n,1+2\gamma;2\lambda
r)\pm \\
\pm nF(1-n,1+2\gamma;2\lambda r)\bigg),
\end{split}
\lab{wf001}
\end{equation} where $\lambda =\sqrt{M^2 -E^2}$
with the eigenvalues given as
\be
E_{nj} =\frac{M}{\sqrt{1+\frac{\alpha^2}{(n+\gamma)^2}}},\;\;\; \gamma =\sqrt{j^2 -\alpha^2},
\ee
where $n$ and $j$ are the principal and angular quantum numbers, respectively.

\begin{figure}[t!]
\centering
\includegraphics[totalheight=0.26\textheight]{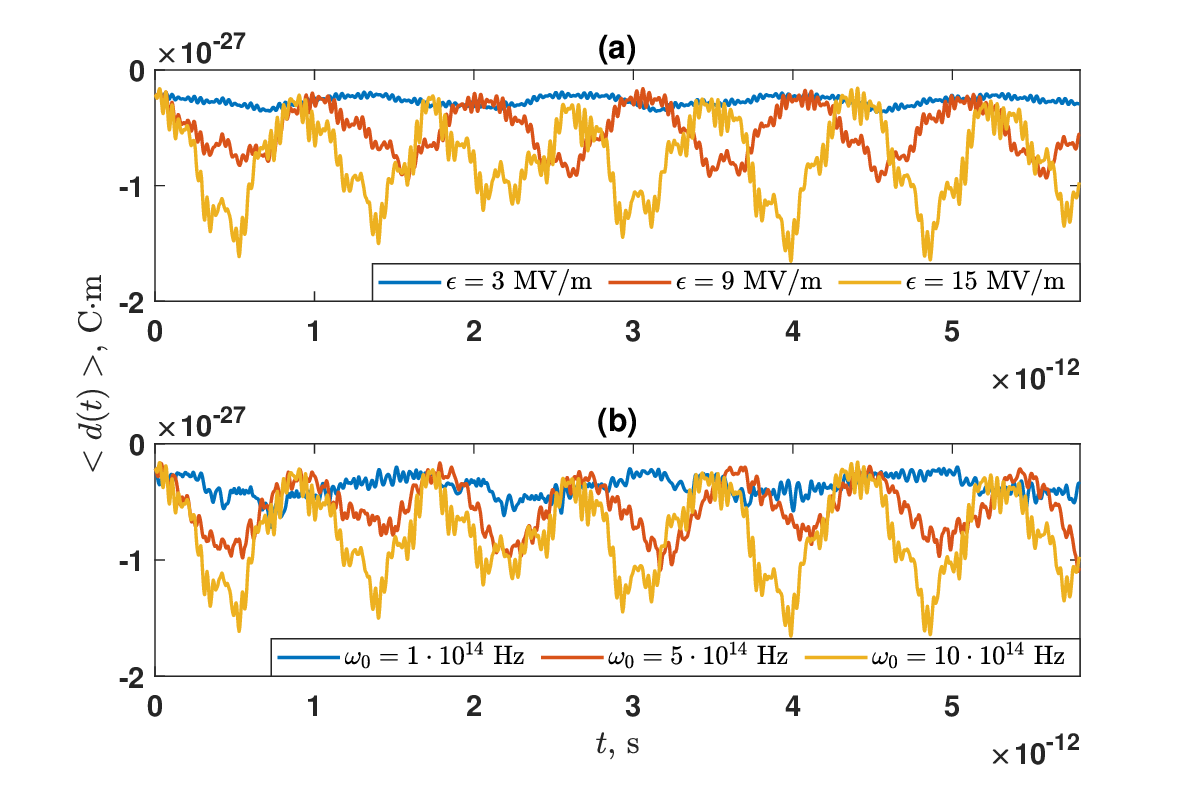}
 \caption{The time dependence of the average dipole moment of the 1D atom in graphene for different values of external field strength for $\omega=10\cdot 10^{14}$ Hz (a) and different values of external field frequency for $\epsilon=15$ MV/m (b) at the value of parameter $\alpha=\frac{300}{137}$, $M\approx0.1$ eV and $v_F=c/300$.}  \label{fig1}
\end{figure}

\section{Optically driven Dirac materials}

Consider interaction of the one-dimensional relativistic atom formed by Coulomb impurity in graphene nanoribbon, with an external, linearly polarized monochromatic field given by
\begin{equation}
    V(x,t)=x\epsilon\cos\omega_0t,
\end{equation}
where  $\epsilon$ is the field strength,   $\omega_0$ is the frequency.

The atom+field system can be described in terms of time-dependent one-dimensional Dirac equation given as
\begin{equation}
    i\frac{\partial\Psi}{\partial t}=\left(\hat{H}_0+V\right)\Psi, \label{eq03}
\end{equation}
where $\hat{H}_0$ is the Dirac operator for unperturbed  one-dimensional hydrogen-like atom.
Solution of Eq.~\re{eq03} can be written in terms of the  complete set of eigenfunctions of unperturbed atom as
\begin{equation}
    \Psi=\sum C_n(t)\psi_n \label{sol03}
\end{equation}
with expansion coefficient $C_n(t)$.

\begin{figure}[t!]
\centering
\includegraphics[totalheight=0.26\textheight]{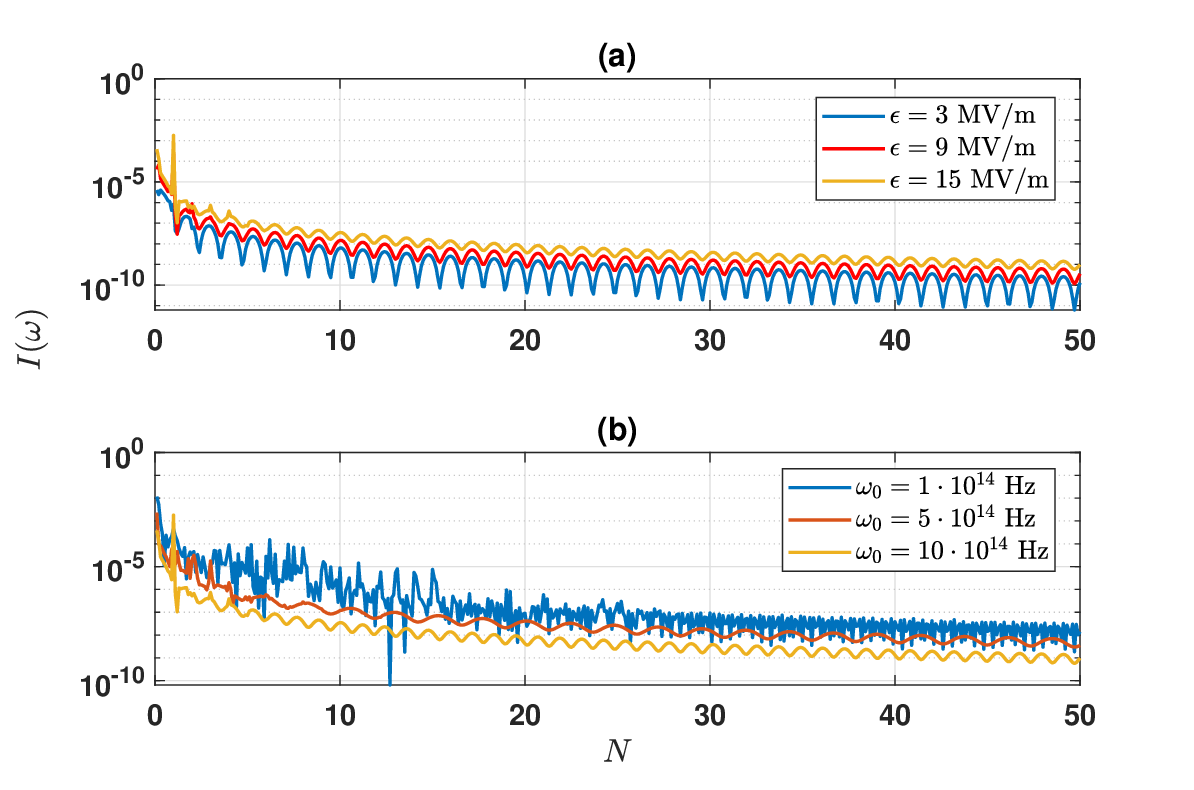}
 \caption{The spectrum of HHG on the 1D atom in graphene for different values of external field strength for $\omega=10\cdot 10^{14}$ Hz (a) and different values of external field frequency for $\epsilon=15$ MV/m (b) at the value of parameter $\alpha=\frac{300}{137}$, $M\approx0.1$ eV and $v_F=c/300$. Note the logarithmic scale for the spectrum of HHG.} \label{fig2}
\end{figure}

Substituting \re{sol03} into the Eq.~\re{eq03}, multiplying $\psi^\dag(x)$ to both side of the equation, integrating with respect to coordinate $x$ and using the orthonormalization condition $\int\psi^\dag_n(x)\psi_m(x)dx=\delta_{nm}$  give us a system of the first order ordinary differential equation as
\begin{equation}
    i\dot{C}_n(t)=C_n(t)E_n+\epsilon\cos\omega_0t\sum_mC_m(t)V_{mn}, \label{eq04}
\end{equation}
where $E_n-$ are the energy levels of the unperturbed atom and the matric element $V_{mn}$ is given by
\begin{equation}
    V_{mn}=\int_0^\infty x\psi_n^\dag(x)\psi_m(x)dx.
    \lab{field}
\end{equation}
Using Eqs. \re{sol01} and \re{sol02} we have
\begin{equation}
\begin{array}{ll}
     V_{mn}=\frac{2C^*_{In}C_{IIm}}{a}\int_0^\infty x\left[ W^{*n}_{\mu,\nu}(\xi)W^{m}_{\mu,\nu}(\xi)+ \right. \\
     \\
     \left. (\kappa_n-iE_n)(\kappa_m+iE_m) W^{*n}_{\mu,\nu+1}(\xi)W^{m}_{\mu,\nu+1}(\xi) \right]dx 
\end{array}
\end{equation}
with $\xi=2\kappa(a+x)$.

The same approach can be used for the treatment of 2D atom with an external optical field given by Eq.\re{field}. Replacing in Eq.\re{sol03} $\psi_n$ with the eigenfunctions of the unperturbed 2D atom given by Eq.\re{wf001} and repeating the same steps as those for 1D atom,  we get 
 the following system of the first order ordinary differential equations: 
\begin{equation}
    i\dot{C}_{nj}(t)=C_{nj}(t)E_{nj}+\epsilon\cos\omega_0t\sum_{n'j'}C_{n'j'}(t)V_{n'nj'j}, \label{eq05}
\end{equation}
where $E_{nj}$ are the eigenvalues of the unperturbed 2D hydrogen-like atom and the matrix element are determined as
\begin{equation}
    V_{n'nj'j}=\int_0^\infty\int_0^{2\pi} \psi_{n'j'}^\dag(r,\theta)r\cos\theta\psi_{nj}(r,\theta)rdrd\theta.
\end{equation}

\begin{figure}[t!]
\centering
\includegraphics[totalheight=0.26\textheight]{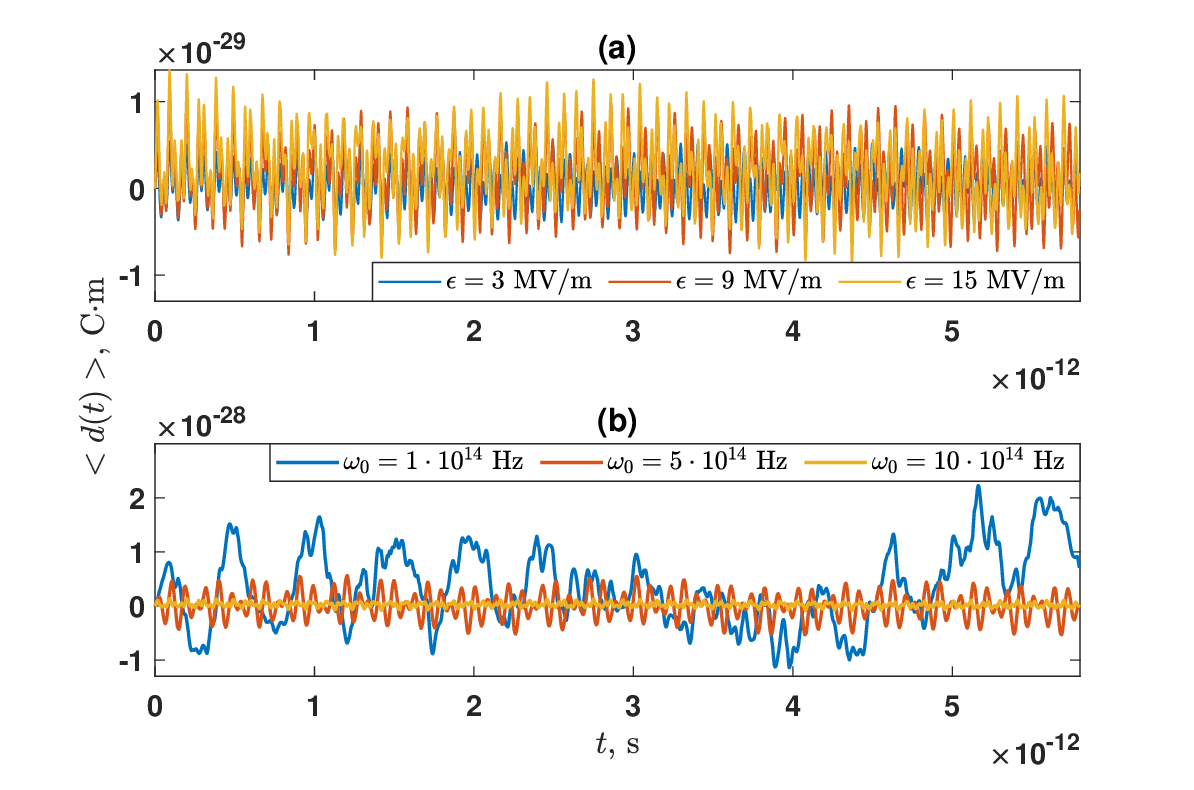}
 \caption{The time dependence of average dipole moment of the electron of the 2D atom in graphene for different values of external field strength for $\omega=10\cdot 10^{14}$ Hz (a) and different values of external field frequency for $\epsilon=15$ MV/m (b) at the value of parameter $\alpha=0.4$, $M\approx0.1$ eV and $v_F=c/300$.}  \label{fig3}
\end{figure}
 
 Using  \re{angwf} and \re{wf001} one can rewrite
\begin{equation}
    V_{n'nj'j}=A_{jj'} \int_0^\infty r^2(F_{n'j'}^\dag F_{nj}+G_{n'j'}^\dag G_{nj})dr, 
\end{equation}
where
\begin{equation}
\begin{split}
    A_{jj'}&=\frac{1}{2\pi}\int_0^{2\pi}\cos\theta e^{i(j-j')\theta}d\theta \\
    &=\left\{ \begin{array}{cc}
         0.5, & \textnormal{if} \hspace{2mm} |j-j'|=1 \\
         0,   & \textnormal{if} \hspace{2mm} |j-j'|\neq 1
    \end{array}. \right.
\end{split}
\end{equation}

Eqs.~\re{eq04} and \re{eq05} are solved numerically to obtain wave functions and for further computations of important characteristics of the "atom + optical field" system. Stability and accuracy of numerical calculations are controlled by means of norm conservation, given as
\begin{equation}
    \sum_n|C_n(t)|^2=1.
\end{equation}

\section{High harmonic generation}
An important phenomenon occurring in the interaction of an optical field with atoms and molecules is so-called optical harmonic generation implying conversion of the frequency of the field due to its interaction with matter. Such effect is a quantum manifestation of nonlinear optical phenomena and can be used for different practical purposes, including ultrashort optical pulse generation and is considered one of the important tools in the attosecond physics \ci{Brabec00,Strelkov16}. The main characteristics of the phenomenon is the intensity of the generated harmonics as a function of frequency: As larger the intensity as acceptable generated frequency for practical use. 
Detailed description of the basic theory of 
high harmonic generation in quantum regime can be found in
\ci{Boydbook}. Here we will consider optical harmonic generation by Coulomb impurities in Dirac materials with focus on the role of relativistic effects. 
Physically important characteristics of
HHG  is the average dipole moment which is determined as
\ci{Boydbook}
\be
\langle d(t) \rangle  = -\langle \Psi(\vec r,t)|x|\Psi(\vec r,t)\rangle .
\ee

 Fig.~\re{fig1}(a) presents the plots of the average dipole moment as a function of time at different values of the external field strength, $\epsilon$ for fixed $\omega=10\cdot 10^{14}$ Hz (a).  Similar plots at different values of external field frequency for fixed $\epsilon=15$ MV/m are presented in Fig.~\re{fig1}(b). For both cases $\alpha$ is fixed as $\alpha=\frac{300}{137}$.  Certain quasiperiodicity  of $\langle d(t)\rangle $ can be seen from the plots.

\begin{figure}[t!]
\centering
\includegraphics[totalheight=0.26\textheight]{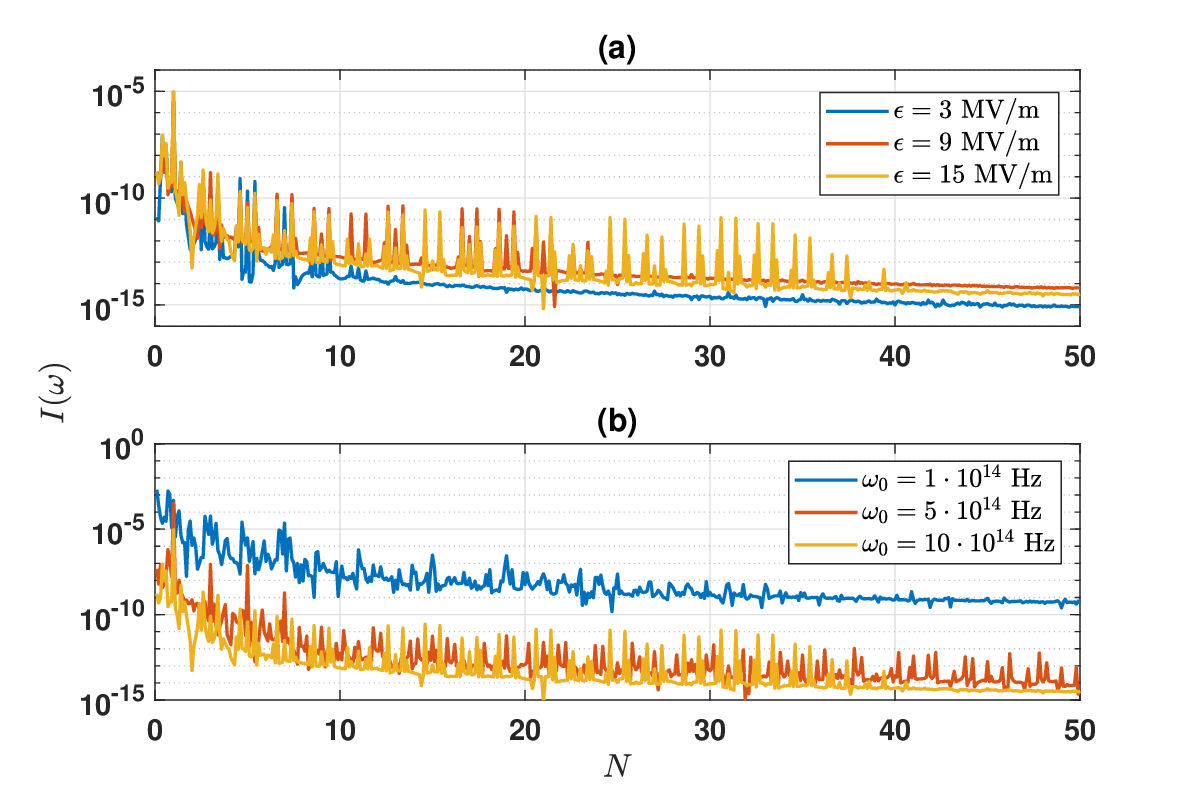}
 \caption{The spectrum of HHG on the 2D atom in graphene for different values of external field strength for $\omega=10\cdot 10^{14}$ Hz (a) and different values of external field frequency for $\epsilon=15$ MV/m (b) at the value of parameter $\alpha=0.4$, $M\approx0.1$ eV and $v_F=c/300$.. Note the logarithmic scale for the spectrum of HHG.} \label{fig4}
\end{figure}

The spectrum of high harmonic generation (HHG) is characterized by the quantity \ci{Boydbook}
\be 
I(\omega)= |\langle  d(\omega)\rangle |^2=\Biggl|\frac{1}{T}\int_{0}^{T} e^{-i\omega t}\langle  d(t) \rangle  dt\Biggl|^2,
\lab{spectr1}
\ee 
where $T$ is the total duration of the interaction.

In Fig.~\re{fig2}(a) optical high harmonic generation spectrum is plotted as a function of harmonic order $N=\omega/\omega_0$  for different values of the external field strength at fixed $\omega=10\cdot 10^{14}$ Hz.  Fig.~\re{fig2}(b) presents similar plots  for different values of external field frequency at fixed $\epsilon=15$ MV/m. In both cases the value of the nucleus charge is chosen as $\alpha=\frac{300}{137}$. Certain, very small plateau can be observed in Fig.~\re{fig2}(b), which is wider as higher the external field strength. 
Relatively slow decay of the intensity starting from the value of the harmonic order, $N=10$ can be seen from the plots. The behavior of the HHG spectrum in Fig.~\re{fig2}(b) shows that its intensity is higher as the external field (fundamental) frequency is smaller. 

Similarly to the above, one can consider HHG in a 2D atom in graphene formed by Coulomb impurities. Fig.~\re{fig3}(a) presents the average dipole moment plotted as a function of time at different values of the external field strength for the fixed  $\omega=10\cdot 10^{14} Hz$. In Fig.~\re{fig3}  plots of $d(t)$ at different values of external field frequency for fixed $\epsilon=15$ MV/m are presented. For both cases the charge of the nucleus is chosen as $\alpha=0.4$. 

In Fig.~\re{fig4}, the spectrum of the HHG in optically driven 2D atom formed by the Coulomb impurity in graphene is presented as a function of harmonic order for different values of the field strength and at fixed value of the fundamental harmonic, $\omega=10\cdot 10^{14}$ Hz. Absence of the plateau and narrow peaks can be seen in Fig.4a.

\section{A sketch for experimental realization}

\begin{figure}[t!]

\subfloat[]{%
  \includegraphics[clip,width=\columnwidth]{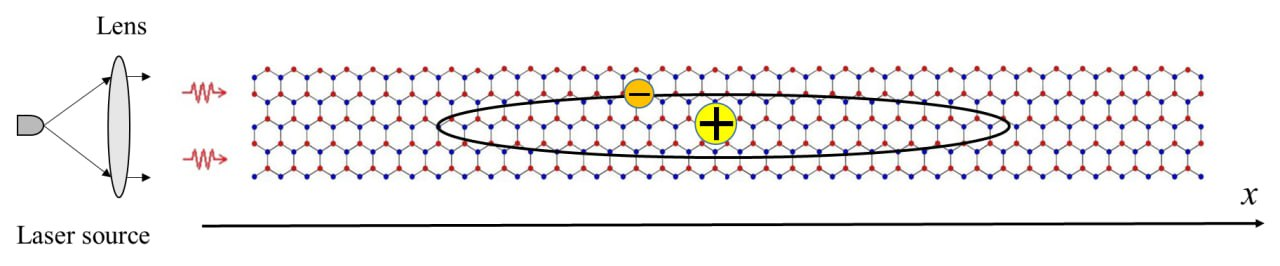}%
}

\subfloat[]{%
  \includegraphics[clip,width=\columnwidth]{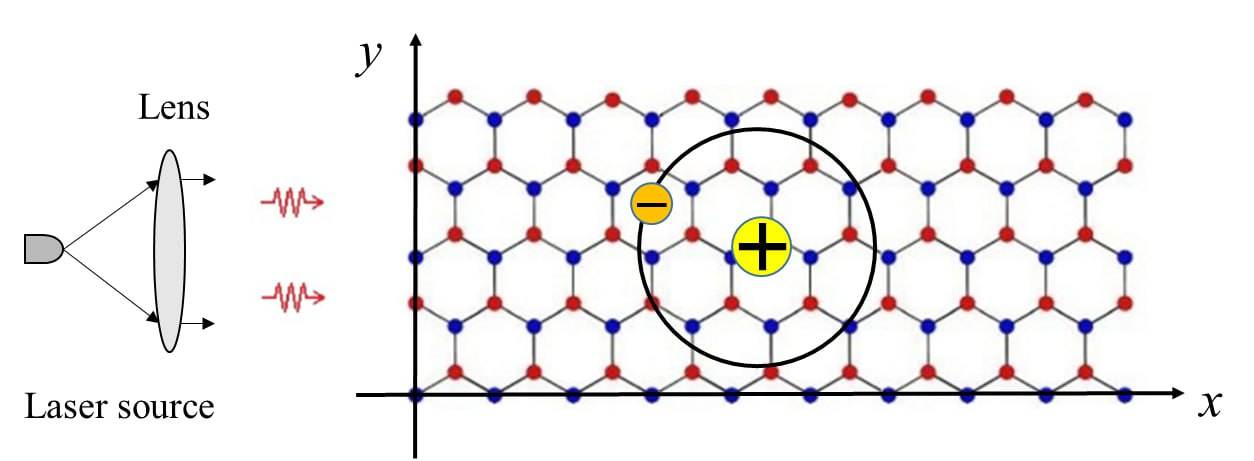}%
}

\caption{Sketch for the experimental realization of a system HHG on the 1D (a) and 2D (b) relativistic atom formed by Coulomb impurity in graphene sheet}

\end{figure}

The above models for high harmonic generation by Coulomb impurities in graphene can be realized in experiment by subjecting bulk graphene sheet (for 2D atom)  and graphene nanoribbon (for 1D atom) to the influence of laser field.

According to the above study, using relatively small power laser should be enough to observe optical high harmonic generation in such interaction. The remarkable feature of the model is caused by the fact that the experimental realization can be done at "table top" level, using small scale experimental set up. Although the above treatment deals with the linearly polarized optical field, the approach we used can be applied for circular and other polarization, while the experimental set up remains the same, except tuning the field polarization.

\section{Conclusion}
In this paper we studied high harmonic generation in relativistic one- and two-dimensional atoms in Dirac materials caused by their interaction with an external optical field. The models are described in terms of the time-dependent Dirac equation with Coulomb and monochromatic optical field potentials. The spectrum of high harmonic generation is calculated at different values of the external field strength and the fundamental harmonic. It is found that in one-dimensional case the HHG intensity has narrow plateau, while for 2D atom only narrow peaks without plateau are observed. A remarkable feature of the above models is caused by the fact that the strength of the external field needed for high harmonic generation is rather small ( few $MV/m$), which is (approximately) by two orders smaller than that of atomic nucleus field strength. Such a feature makes possible experimental realization of the models at table-top level.

\acknowledgements

We acknowledge funding by the grant REP-05032022/235 ("Ultrafast phenomena and vacuum effects in relativistic artificial atoms created in graphene"), funded under the MUNIS Project, supported by the World Bank and the Government of the Republic of Uzbekistan and the grant of the Innovation Development Agency of the Republic of Uzbekistan (Ref. No. FZ-5821512021).

\section*{Author Contribution Statement}
\textbf{S. Rakhmanov and K. Matchonov}: Investigation, Software. \textbf{H. Yusupov and K. Nasriddinov}: Conceptualization, Methodology.  \textbf{D. Matrasulov}: Formal analysis, Writing - original draft.

\section*{Data Availability Statement}
No data was used for the research described in the article.


\begin{thebibliography}{99}

\bibitem{Boydbook} R.W. Boyd, Nonlinear Optics, 3rd ed., Academic Press (2007).
\bibitem{Tong97} X. Tong, Sh. Chu, Chem. Phys.
B, {\bf 217}(2-3), 119, (1997).
\bibitem{Tong05}  J. J. Carrera, Sh. Chu, and X. M. Tong, Phys. Rev. A {\bf 71}, 063813, (2005).
\bibitem{Tong06} X. Guan, X. Tong, and Sh. Chu, Phys. Rev. A {\bf 73}, 023403, (2006).
\bibitem{Milosevic06}  D. B. Milo$\check{\textnormal{s}}$evi\'c,  J. Opt. Soc. Am. B, {\bf 23}, 308, (2006).
\bibitem{Brabec00}  T. Brabec and F. Krausz, Rev. Mod. Phys.,  {\bf 72}(2), 545, (2000).
\bibitem{Yousef06}  I. Yousef and {\it et.al}, Phys. Rep.,  {\bf 427}(2-3), 41, (2006).
\bibitem{Winterfeldt08} C. Winterfeldt, C. Spielmann  and  G. Gerber, Rev. Mod. Phys.,
 {\bf 80}(1), 117, (2008).
\bibitem{Krausz09} F. Krausz, M. Ivanov, Rev. Mod. Phys.,  {\bf 81}, 163, (2009).
\bibitem{Nisoli09} M. Nisoli, G. Sansone, Prog. Quant. Electr., {\bf 33}, 17, (2009).
\bibitem{Kohler12} M.C. Kohler, T. Pfeifer, K.Z. Hatsagortsyan, C.H. Keitel, Advances In Atomic,
Molecular, and Optical Physics, {\bf 61}, 159, (2012).
\bibitem{Strelkov16} V.V. Strelkov, V.T. Platonenko, A.F. Sterzhantov and M.Yu. Ryabikin, Phys. Uspekhi, {\bf 59}(5), 425, (2016).
\bibitem{Rakhmanov20}	S. Rakhmanov, O. V. Karpova, F. S. Khashimova, B. Kh. Eshchanov, Nanosystems: Phys., Chem., Math,, \textbf{11} (3), 307 (2020).
\bibitem{Rakhmanov23}	S.Z. Rakhmanov, I.B. Tursunov, Kh.Sh. Matyokubov, D.U. Matrasulov, Nanosystems: Phys., Chem., Math., \textbf{14}(2), 164, (2023).



\bibitem{Brey06} L. Brey and H. A. Fertig, Phys. Rev. B  {\bf 73}, 235411  (2006).
 \bibitem{Neto09} A.H. Castro Neto, F. Guinea, N.M.R. Perez, K.S. Novoselov, A.K. Geim, Rev. Mod. Phys. \textbf{81}, 109 (2009).
\bibitem{Wu20} J. Wu, Y. Zheng, Z. Zeng, and R. Li, COL \textbf{18}(10), 103201 (2020).
\bibitem{Mrudul21} M. S. Mrudul and G. Dixit, Phys. Rev. B \textbf{103}, 094308 (2021).
\bibitem{Zhang21} Y. Zhang, L. Li, J. Li, T. Huang, P. Lan, and P. Lu, Phys. Rev. A \textbf{104}, 033110 (2021).
\bibitem{Yoshikawa17} N. Yoshikawa, T. Tamaya, K. Tanaka,  Science \textbf{356}, 736 (2017).
\bibitem{Naib14} I. Al-Naib, J. E. Sipe, and M. M. Dignam, Phys. Rev. B \textbf{90}, 245423 (2014).
\bibitem{Sedrakiana23} Kh. V. Sedrakiana, A. G. Ghazaryana, B. R. Avchyana, G. A. Musayelyana, and T. M. Markosyan, JETP, \textbf{137}(1), 47 (2023).
\bibitem{Sorngard13} S. A. Sorngard, S. I. Simonsen, and J. P. Hansen, Phys. Rev. A \textbf{87}, 053803 (2013).
\bibitem{Gnawali23} S. Gnawali and V. Apalkov, Phys. Rev. B \textbf{108}, 115434 (2023).
\bibitem{Gnawali22} S. Gnawali, R. Ghimire, K. R. Magar, S. J. Hossaini, and V. Apalkov, Phys. Rev. B \textbf{106}, 075149 (2022).
\bibitem{Zurron18} O. Zurron, A. Picon and L. Plaja, New J. Phys. \textbf{20} 053033 (2018).
\bibitem{Allor} D. Allor, T. D. Cohen, and D. A. McGady, Phys. Rev. D.  {\bf 78}, 096009, (2008).
\bibitem{Lewk1} M. Lewkowicz, H.C. Kao and B. Rosenstein, Phys. Rev. B.  {\bf 84}, 035414, (2011).
\bibitem{Lewkowicz01} M. Lewkowicz and B. Rosenstein, Phys. Rev. Lett. {\bf 102}, 106802 (2009).

\bibitem{Most1} G. L. Klimchitskaya and V. M. Mostepanenko, Phys. Rev. D \textbf{87}, 125011 (2013).

\bibitem{Fillion} F. Fillion-Gourdeau, and S. MacLean, Phys. Rev. B {\bf 92}, 035401 (2015).


\bibitem{Biswas07} R.R. Biswas, and S. Sachdev, Phys. Rev. B  \textbf{76}, 205122 (2007).
\bibitem{Pereira08} V.M. Pereira, V.N. Kotov, and A. H. Castro Neto, Phys. Rev. B \textbf{78}, 085101 (2008).
\bibitem{Lee13} C.M. Lee, and K.S. Chan, J. Appl. Phys. \textbf{114}, 143708 (2013).
\bibitem{Nishida16} Y. Nishida, Phys. Rev. B  \textbf{94}, 085430 (2016).
\bibitem{Van17} R. Van Pottelberge, M. Zarenia, P. Vasilopoulos, and F. M. Peeters, Phys. Rev. B \textbf{95}, 245410 (2017).


\bibitem{Mayer14} A. Luican-Mayer, M. Kharitonov, G. Li, C.-P. Lu, I. Skachko, A.-M. B. Gon¸calves, K. Watanabe, T. Taniguchi, and E. Y. Andrei, Phys. Rev. Lett. \textbf{112}, 036804 (2014).
\bibitem{Wang13} Y. Wang, D. Wong, A. V. Shytov, V. W. Brar, S. Choi, Q. Wu, H.-Z. Tsai, W. Regan, A. Zettl, R. K. Kawakami, S. G. Louie, L. S. Levitov, and M. F. Crommie, Science 340, 734 (2013).
\bibitem{Wang12} Y. Wang, V. W. Brar, A. V. Shytov, Q. Wu, W. Regan, H.-Z. Tsai, A. Zettl, L. S. Levitov, and M. F. Crommie, Nature Physics \textbf{8}, 653 (2012).
\bibitem{Pereira07} V. M. Pereira, J. Nilsson, and A. H. Castro Neto, Phys. Rev. Lett. \textbf{99}, 166802 (2007).
\bibitem{Gamayun09} O. V. Gamayun, E. V. Gorbar, and V. P. Gusynin, Phys. Rev. B \textbf{80}, 165429 (2009).
\bibitem{Martino14} A. De Martino, D. Kl\"{o}pfer, D. Matrasulov, and R. Egger, Phys. Rev. Lett. \textbf{112}, 186603 (2014).
\bibitem{Klopfer13} D. Kl\"{o}pfer, A. De Martino, and R. Egger, Crystals \textbf{3}, 14 (2013).
\bibitem{Rakhmanov24_01} S. Rakhmanov, R. Egger, and D. Matrasulov, Phys. Scr. \textbf{99}, 075953 (2024).

\bibitem{Novikov} D. Novikov, Phys. Rev. B  {\bf 76}, 245435, (2007).
\bibitem{Khalilov} V.R. Khalilov and Choon-Lin Ho, Mod. Phys.Lett. A  {\bf 13}, 615, (1998).

\bibitem{Downing14} C. A. Downing and M. E. Portnoi, Phys. Rev. A, {\bf 90} 052166 (2014).

\bibitem{Abramowitz} M. Abramowitz, I.A. Stegun, Handbook of Mathematical Functions,
National Bureau of Standards, Washington (1964).

\end{thebibliography}
\end{document}